\documentclass{PoS} 
\usepackage{amssymb,isotope,bm}
\usepackage{amsmath}
\usepackage[export]{adjustbox}
\usepackage[thinlines]{easytable}
\usepackage{mathrsfs}
\usepackage{mathalfa}
\newcommand{\bce}{\begin{center}} \newcommand{\ece}{\end{center}}
\newcommand{\be}{\begin{equation}} \newcommand{\ee}{\end{equation}}
\newcommand{\bea}{\begin{eqnarray}}
\newcommand{\eea}{\end{eqnarray}}
\def\be{\begin{eqnarray}}
\def\ee{\end{eqnarray}}

\newcommand{\vecp}{{\bm p}}
\newcommand{\vecq}{{\bm q}}
\newcommand{\vecv}{{\bm V}}
\newcommand{\vecE}{{\bm E}}
\newcommand{\vece}{{\bm e}}

\newcommand{\vecH}{{\bm H}}
\newcommand{\vech}{{\bm h}}

\newcommand{\vecXi}{{\bm \Xi}}

\def\apj{ApJ}%
\def\aap{A\&A}%
\def\mnras{MNRAS}%
\def\pra{Phys. Rev. A}%
\def\prc{Phys. Rev. C}%
\def\prd{Phys. Rev. D}%
\def\sovast{Soviet Ast.}%
\title{Electrical conductivity tensor of dense plasma in magnetic fields}

\ShortTitle{Electrical conductivity tensor}

\author{\speaker{Arus Harutyunyan}\\
Institute for Theoretical Physics, J.-W. Goethe University,
Frankfurt am Main 
 \email{arus@th.physik.uni-frankfurt.de} 
}

\author{ Armen Sedrakian\\
 Institute for Theoretical Physics, J.-W. Goethe University,
 Frankfurt am Main 
 \email{sedrakian@th.physik.uni-frankfurt.de} 

}

\abstract{ Electrical conductivity of finite-temperature plasma in
  neutron star crusts is studied for applications in
  magneto-hydrodynamical description of compact stars.  We solve the
  Boltzmann kinetic equation in relaxation time approximation taking
  into account the anisotropy of transport due to the magnetic field,
  the effects of dynamical screening in the scattering matrix element
  and correlations among the nuclei.  We show that conductivity has a
  minimum at a non-zero temperature,  a low-temperature 
  decrease and a power-law increase with increasing temperature.
   Selected numerical results are shown for matter composed of carbon,
   iron, and heavier nuclei present in the outer crusts of neutron star. }

\FullConference{The Modern Physics of Compact Stars 2015\\
		30 September 2015 - 3 October 2015\\
		Yerevan, Armenia}

\begin{document}

\section{Introduction}

Magneto-hydrodynamics (MHD) forms the basis of large-scale
description of physics of dense plasma in compact stars.  A key quantity in the
dissipative formulations of MHD is the conductivity of matter. It
determines, for example, the dissipation of currents and therefore the
decay of magnetic fields, the dispersion of plasma waves, etc.  In
turn, magnetic field decay affects the rotational and thermal
evolutions of neutron stars and consequently a broad array of their
observational manifestations.

Transport in compact star plasma was studied traditionally in the cold
(essentially zero-tem\-pe\-rature) and dense regime where the
constituents form degenerate quantum liquids. This regime is relevant
for mature isolated or accreting neutron stars as well as interiors
of white dwarfs. The dilute and warm (non-zero temperature) regime is
of interest in the context of transient, short-lived states of neutron
stars, such as proto-neutron stars newly born in supernova explosions
or hypermassive remnants formed in the aftermath of neutron star
binary mergers.

We start this article with an overview of the transport calculations
of electrical conductivity of compact star matter in the density
regime corresponding to their outer crusts ($\rho \le 10^{11}$ g
cm$^{-3}$). Then we go on to describe our recent effort to calculate
the electrical conductivity of non-zero temperature crustal plasma.
We focus on sufficiently high temperatures where nuclei form a liquid
coexisting with electronic background of arbitrary degeneracy. We
close this review with a summary and outlook.  Below we use the natural (Gaussian) units with
$\hbar= c = k_B = k_e = 1$, $e=\sqrt{\alpha}$, 
$\alpha=1/137$ and the metric signature
$(1,-1,-1,-1)$.

\section{Overview}

\begin{figure}[t] 
\begin{center}
\includegraphics[width=8.0cm,keepaspectratio]{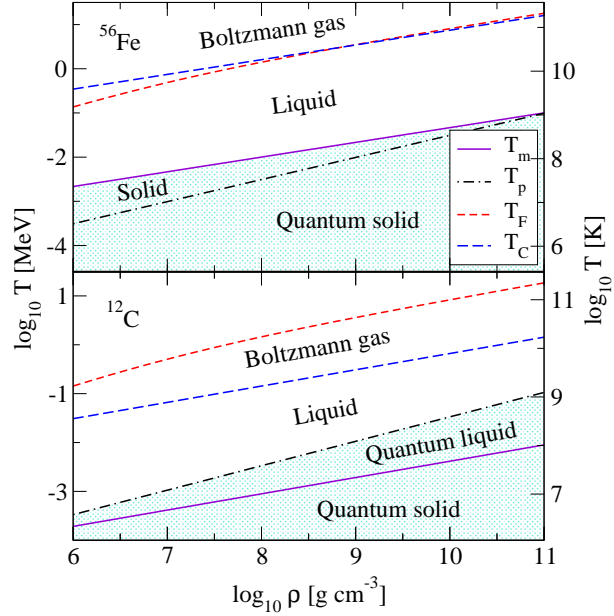}
\caption{ The temperature-density phase diagram of dense plasma 
  composed of iron $\isotope[56]{Fe}$ (upper panel) and carbon 
  $\isotope[12]{C}$ (lower panel). The electron gas degeneracy sets in 
  below the Fermi temperature $T_F$ (short dashed lines). The ionic 
  component solidifies below the melting temperature $T_m$ (solid 
  lines), while quantum effects become important below the plasma 
  temperature (dash-dotted lines).  For temperatures above $T_{\rm C}$
  (long dashed lines) the ionic component forms a Boltzmann gas. Note 
  that for $\isotope[12]{C}$ the quantum effects become important 
  in the portion of the phase diagram lying between the lines 
  $T_p(\rho)$ and $T_m(\rho)$. The present study does not cover the 
  shaded portion of the phase diagram. }
\label{fig:PhaseDiagram} 
\end{center}
\end{figure}

At densities relevant to interiors of white dwarfs and neutron star
crusts the electron-ion system is in a plasma state - the ions are
fully ionized while free electrons are the most mobile carriers of
charge. By charge conservation electron density is related to the ion charge $Z$ 
 by $n_e=Zn_i$, where $n_i$ is the number density of
nuclei. Electrons to a good accuracy form non-interacting gas which
becomes degenerate below the Fermi temperature
$T_F = \varepsilon_F-m = (p_F^2+m^2)^{1/2}-m$, where the electron Fermi
momentum is given by $p_F = (3\pi^2n_e)^{1/3}$ and $m$ is the
electron mass.  The state of ions (mass number $A$ and charge $Z$) is
controlled by the value of the Coulomb plasma parameter $\Gamma$
\bea\label{eq:Gamma}
\Gamma=\frac{e^2 Z^2}{Ta_i}\approx 22.73
\frac{Z^2}{T_6}\bigg(\frac{\rho_6}{A}\bigg)^{1/3},
\eea
where $e$ is the elementary charge, $T$ is the temperature,
$a_i=(4\pi n_i/3)^{-1/3}$ is the radius of the spherical volume per
ion,  $T_6$ is the temperature in
units $10^6$ K and $\rho_6$ is the density in units of
$10^6$  g cm$^{-3}$.  If $\Gamma\ll 1$ or, equivalently
$T\gg T_{\rm C}\equiv Z^2e^2/a_i$, ions form weakly coupled Boltzmann gas.
In the regime $\Gamma\ge 1$ ions are strongly coupled and form a
liquid for low values of $\Gamma\leq\Gamma_m\simeq 160$ and a lattice
for  $\Gamma>\Gamma_m$. The melting temperature of the
lattice associated with $\Gamma_m$ is defined as
$T_m=(Ze)^2/\Gamma_ma_i$.  For
temperatures below the ion plasma temperature
\bea 
T_p = \biggl(\frac{4\pi  Z^2e^2n_i}{M }\biggl)^{1/2} ,
\eea
where $M $ is the ion mass,  the quantization of oscillations of the
lattice becomes important. Figure \ref{fig:PhaseDiagram} shows the
temperature-density phase diagram of the crustal material in the cases
where it is composed of iron $\isotope[56]{Fe}$ (left panel) or carbon
$\isotope[12]{C}$ (right panel).
\begin{figure}[t] 
\begin{center}
\includegraphics[width=8.0cm,keepaspectratio]{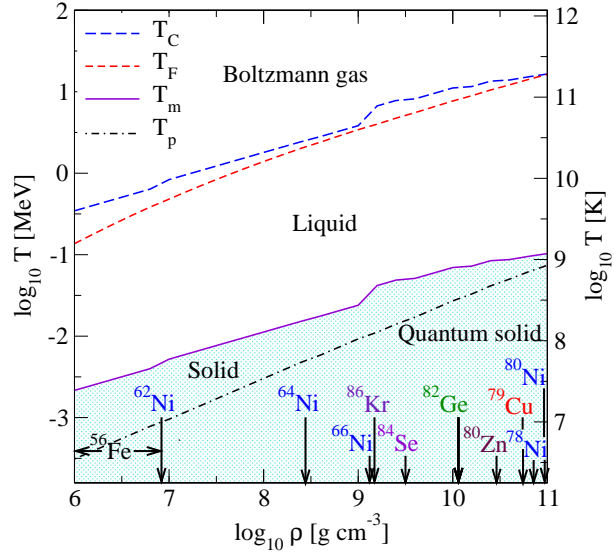}
\caption{ Same as in Fig.~1 but for density
  dependent composition; for details and references see the discussion in Ref.~[12].
}
\label{fig:PhaseDiagram2} 
\end{center}
\end{figure}

\hskip 0.5cm

While the structure of the phase diagrams for $\isotope[56]{Fe}$ and
$\isotope[12]{C}$ are similar there is an important difference as
well: as the temperature is lowered the quantum effects become
important for carbon prior to solidification, whereas iron solidifies
close to the temperature where ionic quantum effects become
important. Except of hydrogen and perhaps helium which may not
solidify because of quantum zero point motions all heavier elements
$Z> 2$ solidify at low enough temperature. The phase diagram in the
case of density dependent composition is shown in
Fig.~\ref{fig:PhaseDiagram2}.

The earliest studies of transport in dense matter go back to the work
by Mestel, Hoyle \cite{1950PCPS...46..331M} and
Lee~\cite{1950ApJ...111..625L} in 1950s, who obtained the ``conductive
opacity'', or equivalently the thermal conductivity of the
electron-ion plasma in non-relativistic electron regime in the context
of radiative and thermal transport in white dwarfs.  Above densities
of the order of $10^6$ g cm$^{-3}$ electrons are
relativistic. Following the initial qualitative estimates of the
conductivity of highly compressed matter by Abrikosov in
1963~\cite{1964Abrikosov} more detailed calculations were carried out
in the 1970s by many authors.  In particular the transport in neutron
star crusts in the relativistic electron regime was studied in much
detail by Flowers and Itoh~\cite{1976ApJ...206..218F} both in the
solid and in the liquid regime using a variational method. They were
able to cover a broad range of densities and low-temperature regime
including multiple channels of scattering, relatively accurate
description of collective modes (phonons) which contribute to the
transport coefficients in the solid phase. Their discussion also
extended to the neutron drip region where free neutrons contribute to
the thermal conductivity and shear viscosity of matter. A critical
analysis of the numerical values for the transport coefficients found
by different authors was given by Yakovlev and
Urpin~\cite{1980SvA....24..303Y}, who also provided useful and simple
approximations for the transport coefficients in the degenerate
electron regime in terms of the Coulomb logarithm.  Nandkumar and
Pethick~\cite{1984MNRAS.209..511N} studied the temperature regime
above the melting temperature, i.e., where ions form a liquid, showing
that the screening of electron-ion interactions can lead to
substantial corrections in this case.  These calculations agree with
those of Itoh et al.~\cite{1983ApJ...273..774I} who also provide
useful fitting formulae for the transport coefficient. Subsequent
refinements of the results quoted above included, among other things,
multi-phonon process and Debye-Waller
factor~\cite{1999A&A...346..345P} in the solid phase and improved
correlation functions in the liquid phase~\cite{2008ApJ...677..495I}.
The implementation of the transport coefficients of dense matter in
the dissipative MHD equations in the case of cold neutron crust plasma
in the presence of magnetic fields were discussed by a number of
authors \cite{1979ApJ...227..995E,1987Ap.....26..295S}. We do not discuss
here the physics in ultra-strong fields and confine our attention to
non-quantizing fields, i.e., fields below the critical field
$B\simeq 10^{14}$~G above which the Landau quantization of electron
trajectories becomes important.

\section{Formalism}
The Boltzmann equation for electron distribution function is given by
\bea\label{eq:Boltzmann}
\frac{\partial f}{\partial t}+
\vecv\frac{\partial f}{\partial\bm r}-
e(\bm E+\vecv \times \bm H)\frac{\partial f}
{\partial\vecp}=I[f],
\eea
where $\vecv$ is the electron velocity, $\vecE$ and $\vecH$ are the
electric and magnetic fields and $I$ is the collision intergal. We are
interested in the regime where the collision integral describes electron-ion
scattering and, therefore, has the form
\bea\label{eq:collision}
I[f]=-(2\pi)^4\sum\limits_{234}|M_{12\to 34}|^2
\delta^{(4)}(p+p_2-p_3-p_4)
[f(1-f_3)g_2-f_3(1-f)g_4],
\eea
where $p_i$ are four-momenta of particles, $g$ is the equilibrium
distribution function of ions, which to a good accuracy can be
described by the Maxwell-Boltzmann distribution with energy spectrum
$\varepsilon=p^2/2M$, where $M$ is the ion mass.  The sum
Eq.~\eqref{eq:collision} stands symbolically for the integrals over
the phase-space of scattering particles, $M_{12\to 34}$ is the
transition matrix element for scattering of relativistic electrons off
correlated ions and is given by
\bea\label{eq:matrix_element}
M_{12\to 34}=\frac{J_0J'_0}{q^2+\Pi_l(\omega,\vecq)}-
\frac{\bm J_t\bm J'_t}{q^2-\omega^2+\Pi_t(\omega,\vecq)},
\eea
where the electron and ion four-currents are given, respectively,
\bea\label{eq:currents}
J^{\mu}=-e^*\bar{u}^{s_3}(p_3)\gamma^\mu u^s(p),\quad
J'^{\mu}=Ze^*V'^{\mu}=Ze^*(1,\vecp'/M),
\eea
$e^* = \sqrt{4\pi}e$, and $J_t, J'_t$ are
the components of the currents transversal to the
moment transfer $\vecq$, $\Pi_l(\omega,\vecq)$ and
$\Pi_t(\omega,\vecq)$ are the longitudinal and transverse components
of the polarization tensor, which describe respectively, the
(irreducible) self-energies of longitudinal and transverse photons in
the medium (plasma).  The form of the matrix element
\eqref{eq:matrix_element} includes thus the dynamical screening of the
electron-ion interaction due to the exchange of transverse
photons. Such separation has been employed in the treatment of
transport in unpaired \cite{1993PhRvD..48.2916H} and superconducting
quark matter~\cite{2014PhRvC..90e5205A} and we adopt an analogous
approach here. We linearize the Boltzmann equation
\eqref{eq:Boltzmann} by writing
\bea\label{eq:perturbation} 
f= f^0+\delta f,\quad \delta f=-\Phi
\frac{\partial f^0}{\partial\varepsilon},  
\eea
where $f^0$ is the equilibrium Fermi-Dirac distribution function,
$\delta f \ll f^0,$ and $\Phi$ is the perturbation.  The electric
field appears in the drift term of linearized Boltzmann equation at
$O(1)$ in perturbation, whereas the term involving magnetic field at
order $O(\Phi)$, because
$
[\vecv\times \vecH]({\partial f^0}/{\partial\vecp})\propto
[\vecv\times \vecH]\vecv=0.
$
We next specify the form of the function $\Phi$ in the case of
conduction as 
$
\Phi=\vecp\cdot \vecXi(\varepsilon), 
$
which after substitution in the linearized Boltzmann equation gives
\bea\label{eq:boltzmann2}
e\vecv\cdot \left[\vecE+(\vecXi\times\vecH)\right]=-
\vecXi\cdot \vecp~\tau^{-1}(\varepsilon),
\eea
where the relaxation time, which depends on electron energy 
$\varepsilon$, is defined by 
\bea\label{eq:t_relax}
\tau^{-1}(\varepsilon)=(2\pi)^{-5}
\int d\omega d\bm q\int d\bm p_2|{M}_{12\to 34}|^2 \frac{\bm q\cdot \bm p}{p^2}
\delta(\varepsilon-\varepsilon_3-\omega)\delta(\varepsilon_2-\varepsilon_4+\omega) 
 g_2\frac{1-f^0_3}{1-f^0}.
\eea
(Here and below the indices 2 and 4 are reserved for ions, the index 3
corresponds to the outgoing electron). 
In transforming the linearized collision integral we introduced a
dummy integration over energy and momentum transfers, i.e., 
$\omega = \varepsilon-\varepsilon_3$ and $\vecq = \vecp-\vecp_3$.  It
remains to express the vector $\vecXi$ describing the perturbation in
terms of physical fields; its most general decomposition is given by
\bea\label{eq:Xi}
\vecXi=\alpha\vece+\beta\vech+\gamma[\vece\times\vech],
\eea
where $\vech \equiv \vecH/H$ and $\vece \equiv \vecE/E$ and the
coefficients $\alpha$, $\beta$, $\gamma$ are functions of the electron
energy. Substituting Eq.~\eqref{eq:Xi} in Eq.~\eqref{eq:boltzmann2}
one finds that $\alpha=-eE\tau /\varepsilon (1+\omega^2_c\tau^2)$,
$\beta/\alpha=(\omega_c\tau)^2(\vece\cdot \vech)$ and
$\gamma/\alpha=-\omega_c\tau$, where $\omega_c=eH\varepsilon^{-1}$ is
the cyclotron frequency. As a result, the most general form of the
perturbation is given by
\bea\label{eq:phi}
\Phi=-\frac{e\tau}{1+(\omega_c\tau)^2}
V_i\left[\delta_{ij}-\omega_c\tau\varepsilon_{ijk}
h_k+(\omega_c\tau)^2h_ih_j\right]E_j.
\eea
Using the standard expression for the 
electrical current in terms of the perturbation $\Phi $ we arrive at
the conductivity tensor 
$
\sigma_{ij}=\delta_{ij}\sigma_0-\varepsilon_{ijm}h_m
\sigma_1 +h_ih_j\sigma_2,
$
where the components of the tensor are  defined as 
\bea\label{eq:sigma2}
\sigma_n=\frac{e^2}{3\pi^2 T}\int_m^\infty\!\! d\varepsilon
\frac{p^3}{\varepsilon}\frac{\tau(\omega_c\tau)^n}
{1+(\omega_c\tau)^2}f^0(1-f^0),\quad n=0,1,2, 
\eea
where $T$ is the temperature and the lower bound of the integral is
given by the mass of the electron, which vanishes in the
ultra-relativistic limit.  The conductivity tensor has a particularly
simple form if the magnetic field is along the $z$-direction
\bea\label{eq:sigma3}
\hat{\sigma}=
\begin{pmatrix}
    \sigma_0 & -\sigma_1 & 0 \\
    \sigma_1 & \sigma_0 & 0 \\
    0 & 0 & \sigma
\end{pmatrix}.
\eea
For zero magnetic field  the current is along the electric field and
we find the scalar conductivity 
\bea\label{eq:sigma}
&&\sigma=\frac{e^2}{3\pi^2 T}\int_m^\infty d\varepsilon
\frac{p^3}{\varepsilon}\tau f^0(1-f^0)=\sigma_0+\sigma_2.
\eea
Thus, the components of the conductivity tensor are fully determined
if the relaxation time $\tau$ is known. We evaluate the square of the
scattering matrix using the standard QFT methods and then average over
the positions of correlated ions, which effectively multiplies the
transition probability by the structure function of ions. After some
computations we find for the relaxation time
\bea\label{eq:relax_time2}
\tau^{-1}(\varepsilon)
&=&\frac{\pi Z^2e^4n_i}{\varepsilon p^3}
\int_{-\infty}^{\varepsilon-m} d\omega e^{-\omega/2T}
\frac{f^0(\varepsilon-\omega)}{f^0(\varepsilon)}
\int_{q_-}^{q_+} dq(q^2-\omega^2+2\varepsilon\omega)S(q)F^2(q) \frac{1}{\sqrt{2\pi}\theta}\nonumber\\
&\times&
e^{-\omega^2/2q^2\theta^2}e^{-q^2/8MT}\bigg\{
\frac{(2\varepsilon-\omega)^2
-q^2}{|q^2+\Pi_l|^2}+\theta^2
\frac{(q^2-\omega^2)[(2\varepsilon-\omega)^2
+q^2]-4m^2q^2}{q^2|q^2-\omega^2+\Pi_t|^2}\bigg\},\nonumber\\
\eea
where $S(q)$ is the ionic structure function, $\theta \equiv \sqrt{{T}/{M}}$, 
$q_{\pm} = \vert\pm p+ \sqrt{p^2-(2\omega\epsilon- \omega^2)}\vert$ and
$\varepsilon = \sqrt{p^2+m^2}$ for non-interacting electrons. The
contribution of longitudinal and transverse photons in
\eqref{eq:relax_time2} separate. The dynamical screening effects
contained in the transverse contribution are parametrically suppressed
by the factor $T/M$ at low temperatures and for heavy nuclei. This
contribution is clearly important in the cases where electron-electron
($e$-$e$) scattering contributes to the collision intergral. This is
the case, for example, when ions form a solid lattice and, therefore,
Umklapp $e$-$e$ processes are allowed, or in the case of thermal conduction
and shear stresses when the $e$-$e$ collisions contribute to the
dissipation. Finally, we note that in order to account for the 
 finite size of the nuclei we have multiplied the transition
 probability in Eq.~\eqref{eq:relax_time2} by the standard expression for 
the nuclear formfactor~\cite{1984ApJ...285..758I}
\bea\label{formfactor}
F(q)=-3\frac{qr_c\cos(qr_c)-\sin(qr_c)}{(qr_c)^3},
\eea
where $r_c$ is the charge radius of the nucleus given by
$r_c=1.15\, A^{1/3}$ fm.

\begin{figure}[t] 
\begin{center}
\includegraphics[width=9cm,height=9cm]{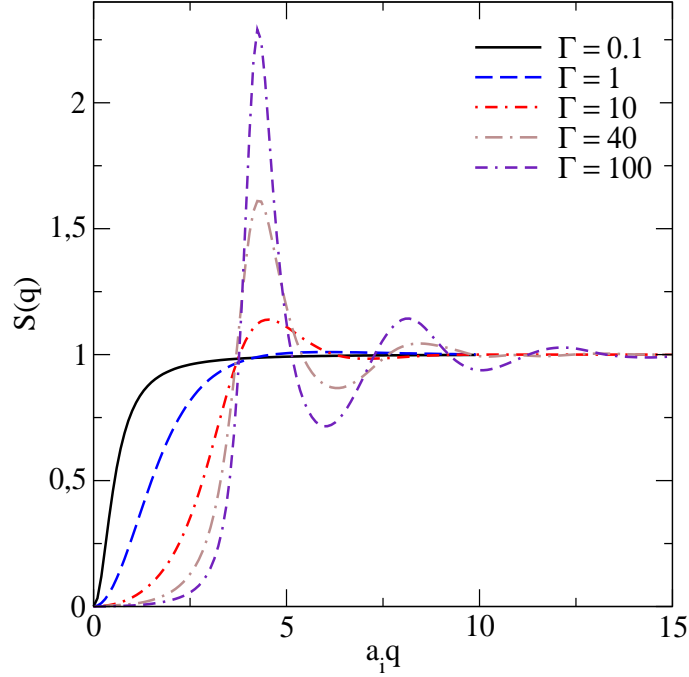}
\caption{ Dependence of the structure function of one-component plasma
  on the magnitude of momentum transfer $q$ in units of inverse $a_i$.
  For $\Gamma\ge 2$ the structure function is taken from Monte-Carlo
  calculations of Galam and Hansen \cite{1976PhRvA..14..816G}. For
  $\Gamma<2$ we obtain the structure function from the analytical
  expressions provided by Tamashiro et al.~\cite{1999PhyA..268...24T}.
}
\label{fig:Sq}
\end{center}
\end{figure}

\section{Results}

For the numerical computations we need to specify the ion structure
function $S(q)$. We assume that only one sort of ions exists at a
given density, so that the structure functions of one-component plasma
(OCP) can be used. These has been extensively computed using various
numerical methods. We adopt the Monte-Carlo results of Galam and
Hansen~\cite{1976PhRvA..14..816G} for Coulomb OCP provided in tabular
form and set a two-dimension spline function in the space spanned by
the magnitude of the momentum transfer $q$ and the plasma parameter
$\Gamma$.  In the low-$\Gamma$ regime ($\Gamma\le 2$) we used the
analytical (leading order) expressions derived by Tamashiro et
al.~\cite{1999PhyA..268...24T} for Coulomb OCP derived using density
functional methods. The resulting structure functions for various
values of the plasma parameter $\Gamma$ are shown in Fig.~\ref{fig:Sq}
as a function of the dimensionless parameter $a_iq$, where $a_i$ is
the ion-radius as defined after Eq.~\eqref{eq:Gamma}.
It is seen that the structure factor universally suppresses
the contribution from small-$q$ scattering. The suppression sets in
for larger $q$ at larger values of $\Gamma$. The large-$q$ asymptotics
is independent of $\Gamma$ as $S(q)\to 1.$ The major difference arises
for intermediate values of $q$ where the structure factor oscillates
and the amplitude of oscillations increases with the value of $\Gamma$
parameter.
The screening of longitudinal and transverse interactions is
determined by the corresponding components of the polarization
tensor. While expression \eqref{eq:relax_time2} is exact with respect
to the form of the polarization tensor, in the numerical calculations
we use the hard-thermal-loop approximation and next-to-leading
expansion in $x = \omega/q$. For the real and imaginary parts of the
polarization tensor we find 
\bea \Pi_l (q,\omega) = q_D^2\chi_l,
\qquad \Pi_t (q,\omega) = q_D^2\chi_t, 
\eea 
where $q_D$ is the Debye
wave-length and the susceptibilities to order $O(x^2)$ are given by
\bea 
\label{eq:chi_l}
&&{\rm Re}\chi_l (q,\omega) = 1-\frac{x^2}{\bar{v}^2},
\quad {\rm Im}\chi_l (q,\omega) =-\frac{\pi x}{2\bar{v}},\\
\label{eq:chi_t}
&&{\rm Re}\chi_t (q,\omega) = x^2,  \qquad {\rm Im}\chi_t
(q,\omega) = \frac{\pi}{4}x\bar{v},
\eea 
where $\bar{v}$ is the electrons average velocity. Because the terms containing $\bar{v}$ are small as well
as electrons are ultra-relativistic  in the most of the regime of interest 
we approximate $\bar{v} =  1$ in our numerical calculations.
For the longitudinal piece of the polarization tensor the screening is
finite in the static case $x = 0 $, while it vanishes for the
transverse piece as $\Pi_t (q,\omega)\propto x^2$, hence the purely
dynamical nature of the transverse screening.
\begin{figure}[t] 
\begin{center}
\includegraphics[width=11cm,height=11cm]{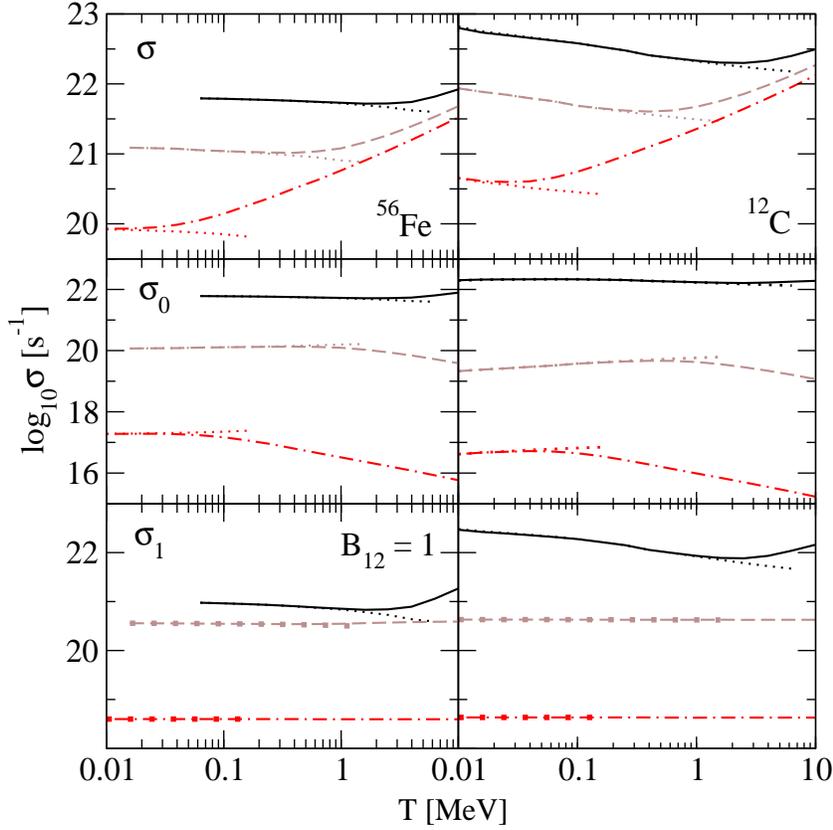}
\caption{ Dependence of the electrical scalar conductivity $\sigma$
  (upper panel) and tensor components $\sigma_0$ (middle panel) and
  $\sigma_1$ (lower panel) on temperature for three densities
  log$_{10}\rho = 10$ (solid lines), log$_{10}\rho = 8$ (dashed lines)
  and log$_{10}\rho = 6$ (dash-dotted lines). The left column contains
  results for $^{56}{\rm Fe}$, the right one for $^{12}{\rm C}$.  The
  dotted lines (symbols in the two lower panels) associated with each
  line show the same, but are evaluate from the zero-temperature Drude
  formula.  The magnetic field is fixed at $B_{12}=1.$}
\label{fig:sigma}
\end{center}
\end{figure}

\begin{figure}[t] 
\begin{center}
\includegraphics[width=11cm,height=11cm]{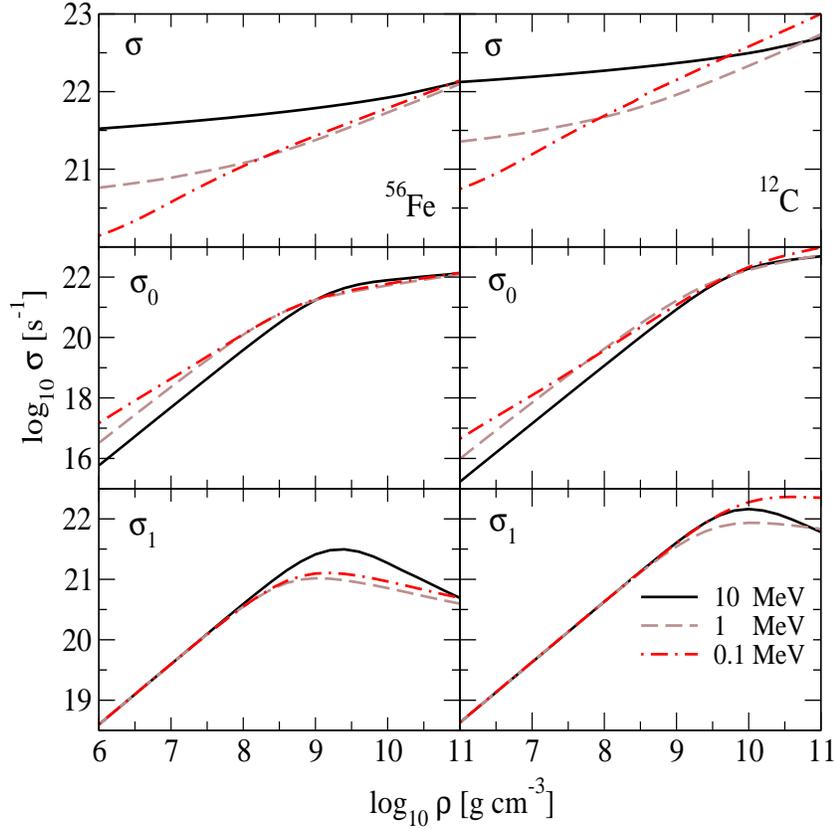}
\caption{Dependence of the electrical scalar conductivity $\sigma$
  (upper panel) and tensor components $\sigma_0$ (middle panel) and 
  $\sigma_1$ (lower panel) on density for three temperatures $T= 10$
  MeV (solid lines), $T= 1$ MeV (dashed lines) and $T= 0.1$ MeV 
  (dash-dotted lines) in the case of $^{56}{\rm Fe}$ (left) and  $C^{12}{\rm 
    C}$ (right) nuclei and   magnetic field $B_{12}=1$. }
\label{fig:sigma2}
\end{center}
\end{figure}

In the zero-temperature limit Eq.~\eqref{eq:sigma} simplifies via the
substitution $T\partial f^0/\partial\varepsilon=- f^0(1-f^0) \to 
-T\delta(\varepsilon-\varepsilon_F)$, i.e.,
\bea\label{sigma_fermi}
\sigma=\frac{e^2}{3\pi^2}\int_m^\infty d\varepsilon
\frac{p^3}{\varepsilon}\tau(\varepsilon)\delta
(\varepsilon-\varepsilon_F)= 
\frac{n_ee^2\tau_F}{\varepsilon_F},
\eea
where $\tau_F$ is the relaxation time (\ref{eq:relax_time2})
taken on the Fermi surface in the $T=0$ limit
\bea\label{eq:relax_fermi}
\tau^{-1}_F\equiv\tau^{-1}(\varepsilon_F)
=\frac{4}{3\pi}Ze^4
\varepsilon_F\int_{0}^{2p_F}dq\frac{q^3}{|q^2+\Pi_l|^2}\bigg(1-\frac{q^2}{4\varepsilon^2_F}\bigg)S(q)F^2(q),
\eea 
where employed the charge neutrality condition $n_e=Zn_i$.  
Neglecting the screening ($\Pi_l\to 0$) 
and the nuclear formfactor [$F(q)\to 1$]
we obtain from \eqref{eq:relax_fermi}
\bea\label{eq:relax_Pethick}
\tau^{-1}_F =\frac{4Ze^4
  \varepsilon_F}{3\pi}\int_{0}^{2p_F}\frac{dq}{q}\bigg(1-\frac{q^2}{4\varepsilon^2_F}\bigg)S(q),
\eea 
which coincides with Eqs. (9) and (11) of
Ref.~\cite{1984MNRAS.209..511N}.

With the input described above we have evaluated the relaxation time
for electron scattering off the ions using Eq.~\eqref{eq:relax_time2}
and then the components of the conductivity tensor according to
Eq.~\eqref{eq:sigma2}. Here we demonstrate selected results, while our
complete results are discussed
elsewhere~\cite{2016arXiv160507612H}. The conductivity as a function
of temperature is shown in Fig.~\ref{fig:sigma} for carbon
$\isotope[12]{C}$ and iron $\isotope[56]{Fe}$ nuclei. The magnitude of
the magnetic field is fixed to $B_{12} = 1$, where $B_{12}$ is the
magnetic field in units of $10^{12}$~G. The full results are compared
to the case where the conductivity is evaluated from the Drude formula
\eqref{sigma_fermi}, which is shown by dotted lines.  The deviation
from the zero temperature result are visible for temperatures in the
range 0.1$-$1 MeV (MeV = $1.16\times 10^{10}$ K) when the density is
varied from $10^6$ to $10^{10}$ g cm$^{-3}$.  It is seen that the
$\sigma$ component of {\it conductivity has a minimum as a function of
temperature:} the low-temperature decrease is replace by a power-law
increase with increasing temperature. This increase can be understood
in terms of the smearing of the Fermi surface by temperature which
makes more electrons available for conduction. The minimum of the
conductivity is one of the key findings of our work. 

The same as in Fig.~\ref{fig:sigma} but as a function of density for
fixed temperature values is shown in Fig.~\ref{fig:sigma2}. The scalar
conductivity and $\sigma_0$ component are increasing functions of
density and depend strongly on the temperature in the low-density
limit, which is associated with lifting of the degeneracy as the
temperature is increased. The behaviour of $\sigma_1$ is reversed: it
is almost independent of temperature and has a maximum.  This is the
consequence of the different scaling of the components of the
conductivity tensor with $\omega_c\tau$ parameter, which describes the
effects of magnetic field. Our results in the cases of matter composed
of $\isotope[12]{C}$ or matter composed of series of nuclei (when the
composition varies with density) show the same general trends as for
$\isotope[56]{Fe}$. The differences between these cases are
quantitative and are discussed in detailed in
Ref.~\cite{2016arXiv160507612H}.

\section{Conclusions}

In this contribution we gave an overview of our current work on the
conductivity of dense matter in the envelopes of neutron
stars at non-zero temperature. One ingredient of our effort is the
formulation of the transport in a manner which allows us to include
the dynamical screening exactly, provided that the polarization tensor
of electrons (or equivalently the self-energies of QED photons) in
plasma can be computed to desired accuracy. Here we employed the
results based on the hard-thermal-loop approximation and low-frequency
expansion appropriate at not very high temperatures. We have shown
that for electron-ion scattering the dynamical screening is suppressed
parametrically by a factor $M/T$, but we anticipate that its effect
would be substantial in the cases (a) of high temperatures and
presence of light clusters, (b) of low temperatures, where the Umklapp
processes with $e$-$e$ scattering are important, (c) of transport
processes where $e$-$e$ scattering may be dominant from the outset,
such as the thermal conductivity and shear viscosity.

Our numerical results show that the scalar conductivity (no anisotropy
due to the $B$-field) has a minimum as a function of temperature, with
a power-law decrease at low-temperatures and a power-law increase at
higher temperatures. The range of validity of the zero-temperature
Drude formula extends from low temperatures up to 0.1$-$1 MeV
($10^9-10^{10}$ K), where the lower of these bounds corresponds to
density $\rho \sim 10^{6}$ g cm$^{-3}$ and the upper one to
$\rho \sim 10^{10}$ g cm$^{-3}$. The behaviour of the off-diagonal
$\sigma_1$ component of the conductivity tensor is similar to the one
described above except at low densities, where it remains almost
constant. Finally the $\sigma_0$ component shows strongly density dependent
behaviour: for large densities (high degeneracy) it behaves analogous
to $\sigma$, but shows the inverse trend for low-densities which is
associated with the transition from the regime $\omega_c\tau <1 $ to
$\omega_c\tau >1 $.

\section*{Acknowledgments}

We thank M. Alford, H. Nishimura, L. Rezzolla and D. Rischke for
discussions. This work was supported by the HGS-HIRe graduate program
at Frankfurt University (A. H.), by the Deutsche
Forschungsgemeinschaft's Grant No. SE 1836/3-1 (A. S.)  and by the
NewCompStar COST Action MP1304.  We thank the Volkswagen Stiftung
for the support of the 2015 edition of the conference series {\it
  ``The Modern Physics of Compact Stars and Relativistic Gravity''}.

\bibliographystyle{JHEP}
\providecommand{\href}[2]{#2}\begingroup\raggedright\endgroup

\section*{Supplemental material}

Below we present numerical tables for the conductivities log$_{10}\sigma$, log$_{10}\sigma_0$ and log$_{10}\sigma_1$ (in units of s$^{-1}$) for various values of magnetic field (in units of $10^{12}$ G) for sets of values of density [g cm$^{-3}$] and  temperature [MeV].  The tables are
 provided for three types of composition of matter: $\isotope[12]{C}$ nuclei, $\isotope[56]{Fe}$ nuclei, and density-dependent composition as indicated in Fig.~2.  Analytical fits to these results with  relative error $\le 10\%$  can be found in Ref.~\cite{2016arXiv160507612H}. 

\begin{table}
\begin{center}
\vskip -1cm 
\caption {$\log\sigma$ for $\isotope[12]{C}$} \label{tab:sigma_C}
{\tiny
\begin{tabular}{ |c||c|c|c|c|c|c|c|c|c|c|c| }
\hline
  $\log\rho$  &  $\log T$=-1.0  &  $\log T$=-0.8  &  $\log T$=-0.6  &  $\log T$=-0.4  &  $\log T$=-0.2  &  $\log T$=0.0  &  $\log T$=0.2  &  $\log T$=0.4  &  $\log T$=0.6  &  $\log T$=0.8  &  $\log T$=1.0  \\
  \hline
6.0  &  20.745  &  20.853  &  20.973  &  21.100  &  21.222  &  21.356  &  21.498  &  21.646  &  21.798  &  21.957  &  22.123   \\
 6.2  &  20.816  &  20.905  &  21.013  &  21.131  &  21.251  &  21.378  &  21.518  &  21.663  &  21.814  &  21.970  &  22.136   \\
 6.4  &  20.898  &  20.965  &  21.057  &  21.165  &  21.281  &  21.401  &  21.538  &  21.681  &  21.830  &  21.985  &  22.149   \\
 6.6  &  20.990  &  21.033  &  21.106  &  21.202  &  21.313  &  21.426  &  21.559  &  21.699  &  21.846  &  22.000  &  22.162   \\
 6.8  &  21.089  &  21.110  &  21.163  &  21.244  &  21.346  &  21.452  &  21.581  &  21.719  &  21.864  &  22.015  &  22.176   \\
 7.0  &  21.192  &  21.193  &  21.226  &  21.290  &  21.381  &  21.482  &  21.605  &  21.739  &  21.882  &  22.031  &  22.190   \\
 7.2  &  21.296  &  21.283  &  21.296  &  21.342  &  21.420  &  21.518  &  21.630  &  21.760  &  21.900  &  22.048  &  22.205   \\
\hline
\end{tabular}
} 
\caption {$\log\sigma_0$ for $\isotope[12]{C}$ at $B_{12}=1$} \label{tab:sigma0_b12_C}
{\tiny
\begin{tabular}{ |c||c|c|c|c|c|c|c|c|c|c|c| }
\hline
  $\log\rho$  &  $\log T$=-1.0  &  $\log T$=-0.8  &  $\log T$=-0.6  &  $\log T$=-0.4  &  $\log T$=-0.2  &  $\log T$=0.0  &  $\log T$=0.2  &  $\log T$=0.4  &  $\log T$=0.6  &  $\log T$=0.8  &  $\log T$=1.0  \\
  \hline
6.0  &  16.648  &  16.551  &  16.423  &  16.279  &  16.132  &  15.987  &  15.843  &  15.696  &  15.546  &  15.390  &  15.227   \\
 6.2  &  16.947  &  16.878  &  16.771  &  16.641  &  16.503  &  16.363  &  16.222  &  16.078  &  15.929  &  15.776  &  15.613   \\
 6.4  &  17.238  &  17.195  &  17.112  &  16.999  &  16.871  &  16.736  &  16.599  &  16.458  &  16.312  &  16.160  &  15.999   \\
 6.6  &  17.522  &  17.504  &  17.445  &  17.352  &  17.236  &  17.108  &  16.976  &  16.838  &  16.694  &  16.544  &  16.385   \\
 6.8  &  17.805  &  17.806  &  17.771  &  17.699  &  17.597  &  17.478  &  17.351  &  17.216  &  17.075  &  16.928  &  16.770   \\
 7.0  &  18.090  &  18.104  &  18.091  &  18.040  &  17.955  &  17.845  &  17.724  &  17.594  &  17.455  &  17.310  &  17.155   \\
 7.2  &  18.378  &  18.402  &  18.405  &  18.375  &  18.307  &  18.211  &  18.096  &  17.970  &  17.835  &  17.692  &  17.539   \\
\hline
\end{tabular}
} 
\caption {$\log\sigma_0$ for $\isotope[12]{C}$ at $B_{12}=10$} \label{tab:sigma0_b13_C}
{\tiny
\begin{tabular}{ |c||c|c|c|c|c|c|c|c|c|c|c| }
\hline
  $\log\rho$  &  $\log T$=-1.0  &  $\log T$=-0.8  &  $\log T$=-0.6  &  $\log T$=-0.4  &  $\log T$=-0.2  &  $\log T$=0.0  &  $\log T$=0.2  &  $\log T$=0.4  &  $\log T$=0.6  &  $\log T$=0.8  &  $\log T$=1.0  \\
  \hline
6.0  &  14.648  &  14.551  &  14.423  &  14.280  &  14.133  &  13.988  &  13.842  &  13.696  &  13.546  &  13.390  &  13.227   \\
 6.2  &  14.948  &  14.878  &  14.771  &  14.642  &  14.503  &  14.363  &  14.221  &  14.078  &  13.929  &  13.776  &  13.613   \\
 6.4  &  15.238  &  15.195  &  15.112  &  14.999  &  14.871  &  14.736  &  14.599  &  14.458  &  14.312  &  14.160  &  13.999   \\
 6.6  &  15.523  &  15.504  &  15.446  &  15.352  &  15.236  &  15.108  &  14.975  &  14.838  &  14.694  &  14.544  &  14.385   \\
 6.8  &  15.806  &  15.806  &  15.772  &  15.700  &  15.598  &  15.478  &  15.351  &  15.216  &  15.075  &  14.928  &  14.770   \\
 7.0  &  16.090  &  16.105  &  16.092  &  16.041  &  15.955  &  15.846  &  15.724  &  15.594  &  15.455  &  15.310  &  15.155   \\
 7.2  &  16.379  &  16.403  &  16.406  &  16.376  &  16.308  &  16.212  &  16.096  &  15.970  &  15.835  &  15.692  &  15.539   \\
\hline
\end{tabular}
} 
\end{center}
\end{table}
\begin{table}
\begin{center}
\vskip -1cm 
\caption {$\log\sigma_0$ for $\isotope[12]{C}$ at $B_{12}=100$} \label{tab:sigma0_b14_C}
{\tiny
\begin{tabular}{ |c||c|c|c|c|c|c|c|c|c|c|c| }
\hline
  $\log\rho$  &  $\log T$=-1.0  &  $\log T$=-0.8  &  $\log T$=-0.6  &  $\log T$=-0.4  &  $\log T$=-0.2  &  $\log T$=0.0  &  $\log T$=0.2  &  $\log T$=0.4  &  $\log T$=0.6  &  $\log T$=0.8  &  $\log T$=1.0  \\
  \hline
6.0  &  12.648  &  12.551  &  12.423  &  12.280  &  12.133  &  11.988  &  11.843  &  11.696  &  11.546  &  11.390  &  11.227   \\
 6.2  &  12.948  &  12.878  &  12.771  &  12.642  &  12.503  &  12.363  &  12.222  &  12.078  &  11.929  &  11.776  &  11.613   \\
 6.4  &  13.238  &  13.195  &  13.112  &  12.999  &  12.871  &  12.736  &  12.599  &  12.458  &  12.312  &  12.160  &  11.999   \\
 6.6  &  13.523  &  13.504  &  13.446  &  13.352  &  13.236  &  13.108  &  12.976  &  12.838  &  12.694  &  12.544  &  12.385   \\
 6.8  &  13.806  &  13.806  &  13.772  &  13.700  &  13.598  &  13.478  &  13.351  &  13.216  &  13.075  &  12.928  &  12.770   \\
 7.0  &  14.090  &  14.105  &  14.092  &  14.041  &  13.955  &  13.846  &  13.724  &  13.594  &  13.455  &  13.310  &  13.155   \\
 7.2  &  14.379  &  14.403  &  14.406  &  14.376  &  14.308  &  14.212  &  14.096  &  13.970  &  13.835  &  13.692  &  13.539   \\
\hline
\end{tabular}
} 
\caption {$\log\sigma_1$ for $\isotope[12]{C}$ at $B_{12}=1$} \label{tab:sigma1_b12_C}
{\tiny
\begin{tabular}{ |c||c|c|c|c|c|c|c|c|c|c|c| }
\hline
  $\log\rho$  &  $\log T$=-1.0  &  $\log T$=-0.8  &  $\log T$=-0.6  &  $\log T$=-0.4  &  $\log T$=-0.2  &  $\log T$=0.0  &  $\log T$=0.2  &  $\log T$=0.4  &  $\log T$=0.6  &  $\log T$=0.8  &  $\log T$=1.0  \\
  \hline
6.0  &  18.633  &  18.633  &  18.633  &  18.633  &  18.629  &  18.629  &  18.629  &  18.629  &  18.629  &  18.629  &  18.629   \\
 6.2  &  18.833  &  18.833  &  18.833  &  18.833  &  18.831  &  18.829  &  18.829  &  18.829  &  18.829  &  18.829  &  18.829   \\
 6.4  &  19.033  &  19.033  &  19.033  &  19.033  &  19.032  &  19.029  &  19.029  &  19.029  &  19.029  &  19.029  &  19.029   \\
 6.6  &  19.233  &  19.233  &  19.233  &  19.233  &  19.233  &  19.229  &  19.229  &  19.229  &  19.229  &  19.229  &  19.229   \\
 6.8  &  19.433  &  19.433  &  19.433  &  19.433  &  19.433  &  19.429  &  19.429  &  19.429  &  19.429  &  19.429  &  19.429   \\
 7.0  &  19.633  &  19.633  &  19.633  &  19.633  &  19.633  &  19.629  &  19.629  &  19.629  &  19.629  &  19.629  &  19.629   \\
 7.2  &  19.833  &  19.833  &  19.833  &  19.833  &  19.833  &  19.831  &  19.829  &  19.829  &  19.829  &  19.829  &  19.829   \\
\hline
\end{tabular}
} 
\caption {$\log\sigma_1$ for $\isotope[12]{C}$ at $B_{12}=10$} \label{tab:sigma1_b13_C}
{\tiny
\begin{tabular}{ |c||c|c|c|c|c|c|c|c|c|c|c| }
\hline
  $\log\rho$  &  $\log T$=-1.0  &  $\log T$=-0.8  &  $\log T$=-0.6  &  $\log T$=-0.4  &  $\log T$=-0.2  &  $\log T$=0.0  &  $\log T$=0.2  &  $\log T$=0.4  &  $\log T$=0.6  &  $\log T$=0.8  &  $\log T$=1.0  \\
  \hline
6.0  &  17.633  &  17.633  &  17.633  &  17.633  &  17.629  &  17.629  &  17.629  &  17.629  &  17.629  &  17.629  &  17.629   \\
 6.2  &  17.833  &  17.833  &  17.833  &  17.833  &  17.831  &  17.829  &  17.829  &  17.829  &  17.829  &  17.829  &  17.829   \\
 6.4  &  18.033  &  18.033  &  18.033  &  18.033  &  18.032  &  18.029  &  18.029  &  18.029  &  18.029  &  18.029  &  18.029   \\
 6.6  &  18.233  &  18.233  &  18.233  &  18.233  &  18.233  &  18.229  &  18.229  &  18.229  &  18.229  &  18.229  &  18.229   \\
 6.8  &  18.433  &  18.433  &  18.433  &  18.433  &  18.433  &  18.429  &  18.429  &  18.429  &  18.429  &  18.429  &  18.429   \\
 7.0  &  18.633  &  18.633  &  18.633  &  18.633  &  18.633  &  18.629  &  18.629  &  18.629  &  18.629  &  18.629  &  18.629   \\
 7.2  &  18.833  &  18.833  &  18.833  &  18.833  &  18.833  &  18.832  &  18.829  &  18.829  &  18.829  &  18.829  &  18.829   \\
\hline
\end{tabular}
} 
\end{center}
\end{table}
\begin{table}
\begin{center}
\vskip -1cm 
\caption {$\log\sigma_1$ for $\isotope[12]{C}$ at $B_{12}=100$} \label{tab:sigma1_b14_C}
{\tiny
\begin{tabular}{ |c||c|c|c|c|c|c|c|c|c|c|c| }
\hline
  $\log\rho$  &  $\log T$=-1.0  &  $\log T$=-0.8  &  $\log T$=-0.6  &  $\log T$=-0.4  &  $\log T$=-0.2  &  $\log T$=0.0  &  $\log T$=0.2  &  $\log T$=0.4  &  $\log T$=0.6  &  $\log T$=0.8  &  $\log T$=1.0  \\
  \hline
6.0  &  16.633  &  16.633  &  16.633  &  16.633  &  16.629  &  16.629  &  16.629  &  16.629  &  16.629  &  16.629  &  16.629   \\
 6.2  &  16.833  &  16.833  &  16.833  &  16.833  &  16.831  &  16.829  &  16.829  &  16.829  &  16.829  &  16.829  &  16.829   \\
 6.4  &  17.033  &  17.033  &  17.033  &  17.033  &  17.032  &  17.029  &  17.029  &  17.029  &  17.029  &  17.029  &  17.029   \\
 6.6  &  17.233  &  17.233  &  17.233  &  17.233  &  17.233  &  17.229  &  17.229  &  17.229  &  17.229  &  17.229  &  17.229   \\
 6.8  &  17.433  &  17.433  &  17.433  &  17.433  &  17.433  &  17.429  &  17.429  &  17.429  &  17.429  &  17.429  &  17.429   \\
 7.0  &  17.633  &  17.633  &  17.633  &  17.633  &  17.633  &  17.629  &  17.629  &  17.629  &  17.629  &  17.629  &  17.629   \\
 7.2  &  17.833  &  17.833  &  17.833  &  17.833  &  17.833  &  17.832  &  17.829  &  17.829  &  17.829  &  17.829  &  17.829   \\
\hline
\end{tabular}
} 
\caption {$\log\sigma$ for $\isotope[56]{Fe}$} \label{tab:sigma_Fe}
{\tiny
\begin{tabular}{ |c||c|c|c|c|c|c|c|c|c|c|c| }
\hline
  $\log\rho$  &  $\log T$=-1.0  &  $\log T$=-0.8  &  $\log T$=-0.6  &  $\log T$=-0.4  &  $\log T$=-0.2  &  $\log T$=0.0  &  $\log T$=0.2  &  $\log T$=0.4  &  $\log T$=0.6  &  $\log T$=0.8  &  $\log T$=1.0  \\
  \hline
6.0  &  20.144  &  20.260  &  20.375  &  20.506  &  20.628  &  20.762  &  20.902  &  21.048  &  21.200  &  21.357  &  21.522   \\
 6.2  &  20.213  &  20.310  &  20.413  &  20.537  &  20.658  &  20.785  &  20.923  &  21.066  &  21.216  &  21.372  &  21.536   \\
 6.4  &  20.293  &  20.368  &  20.456  &  20.570  &  20.689  &  20.809  &  20.944  &  21.085  &  21.233  &  21.388  &  21.550   \\
 6.6  &  20.382  &  20.435  &  20.517  &  20.607  &  20.720  &  20.834  &  20.966  &  21.104  &  21.250  &  21.403  &  21.565   \\
 6.8  &  20.477  &  20.510  &  20.572  &  20.647  &  20.753  &  20.861  &  20.989  &  21.125  &  21.268  &  21.420  &  21.580   \\
 7.0  &  20.576  &  20.592  &  20.634  &  20.692  &  20.788  &  20.889  &  21.013  &  21.146  &  21.287  &  21.437  &  21.595   \\
 7.2  &  20.675  &  20.680  &  20.702  &  20.756  &  20.826  &  20.925  &  21.038  &  21.168  &  21.307  &  21.454  &  21.611   \\
\hline
\end{tabular}
} 
\caption {$\log\sigma_0$ for $\isotope[56]{Fe}$ at $B_{12}=1$} \label{tab:sigma0_b12_Fe}
{\tiny
\begin{tabular}{ |c||c|c|c|c|c|c|c|c|c|c|c| }
\hline
  $\log\rho$  &  $\log T$=-1.0  &  $\log T$=-0.8  &  $\log T$=-0.6  &  $\log T$=-0.4  &  $\log T$=-0.2  &  $\log T$=0.0  &  $\log T$=0.2  &  $\log T$=0.4  &  $\log T$=0.6  &  $\log T$=0.8  &  $\log T$=1.0  \\
  \hline
6.0  &  17.169  &  17.065  &  16.945  &  16.800  &  16.654  &  16.512  &  16.370  &  16.226  &  16.078  &  15.925  &  15.766   \\
 6.2  &  17.472  &  17.395  &  17.295  &  17.161  &  17.023  &  16.886  &  16.748  &  16.607  &  16.460  &  16.309  &  16.151   \\
 6.4  &  17.768  &  17.715  &  17.637  &  17.519  &  17.391  &  17.258  &  17.124  &  16.986  &  16.842  &  16.692  &  16.536   \\
 6.6  &  18.057  &  18.026  &  17.957  &  17.872  &  17.755  &  17.629  &  17.499  &  17.364  &  17.223  &  17.075  &  16.920   \\
 6.8  &  18.345  &  18.331  &  18.285  &  18.219  &  18.115  &  17.997  &  17.873  &  17.742  &  17.603  &  17.457  &  17.304   \\
 7.0  &  18.634  &  18.631  &  18.605  &  18.560  &  18.471  &  18.362  &  18.245  &  18.117  &  17.982  &  17.839  &  17.687   \\
 7.2  &  18.926  &  18.930  &  18.919  &  18.879  &  18.821  &  18.725  &  18.614  &  18.492  &  18.360  &  18.219  &  18.070   \\
\hline
\end{tabular}
} 
\end{center}
\end{table}
\begin{table}
\begin{center}
\vskip -1cm 
\caption {$\log\sigma_0$ for $\isotope[56]{Fe}$ at $B_{12}=10$} \label{tab:sigma0_b13_Fe}
{\tiny
\begin{tabular}{ |c||c|c|c|c|c|c|c|c|c|c|c| }
\hline
  $\log\rho$  &  $\log T$=-1.0  &  $\log T$=-0.8  &  $\log T$=-0.6  &  $\log T$=-0.4  &  $\log T$=-0.2  &  $\log T$=0.0  &  $\log T$=0.2  &  $\log T$=0.4  &  $\log T$=0.6  &  $\log T$=0.8  &  $\log T$=1.0  \\
  \hline
6.0  &  15.171  &  15.068  &  14.948  &  14.801  &  14.655  &  14.512  &  14.370  &  14.226  &  14.078  &  13.925  &  13.766   \\
 6.2  &  15.475  &  15.398  &  15.298  &  15.163  &  15.024  &  14.886  &  14.748  &  14.607  &  14.460  &  14.309  &  14.151   \\
 6.4  &  15.771  &  15.718  &  15.641  &  15.522  &  15.392  &  15.259  &  15.125  &  14.986  &  14.842  &  14.692  &  14.536   \\
 6.6  &  16.061  &  16.031  &  15.961  &  15.875  &  15.757  &  15.630  &  15.500  &  15.365  &  15.223  &  15.075  &  14.920   \\
 6.8  &  16.350  &  16.336  &  16.290  &  16.224  &  16.118  &  15.999  &  15.874  &  15.742  &  15.603  &  15.457  &  15.304   \\
 7.0  &  16.640  &  16.638  &  16.612  &  16.568  &  16.476  &  16.366  &  16.246  &  16.119  &  15.983  &  15.839  &  15.688   \\
 7.2  &  16.934  &  16.939  &  16.929  &  16.889  &  16.830  &  16.731  &  16.617  &  16.494  &  16.361  &  16.220  &  16.070   \\
\hline
\end{tabular}
} 
\caption {$\log\sigma_0$ for $\isotope[56]{Fe}$ at $B_{12}=100$} \label{tab:sigma0_b14_Fe}
{\tiny
\begin{tabular}{ |c||c|c|c|c|c|c|c|c|c|c|c| }
\hline
  $\log\rho$  &  $\log T$=-1.0  &  $\log T$=-0.8  &  $\log T$=-0.6  &  $\log T$=-0.4  &  $\log T$=-0.2  &  $\log T$=0.0  &  $\log T$=0.2  &  $\log T$=0.4  &  $\log T$=0.6  &  $\log T$=0.8  &  $\log T$=1.0  \\
  \hline
6.0  &  13.171  &  13.068  &  12.948  &  12.801  &  12.655  &  12.512  &  12.370  &  12.226  &  12.078  &  11.925  &  11.766   \\
 6.2  &  13.475  &  13.398  &  13.298  &  13.163  &  13.025  &  12.886  &  12.748  &  12.607  &  12.460  &  12.309  &  12.151   \\
 6.4  &  13.771  &  13.718  &  13.641  &  13.522  &  13.392  &  13.259  &  13.125  &  12.986  &  12.842  &  12.692  &  12.536   \\
 6.6  &  14.061  &  14.031  &  13.961  &  13.876  &  13.757  &  13.630  &  13.500  &  13.365  &  13.223  &  13.075  &  12.920   \\
 6.8  &  14.350  &  14.336  &  14.290  &  14.224  &  14.118  &  13.999  &  13.874  &  13.742  &  13.603  &  13.457  &  13.304   \\
 7.0  &  14.640  &  14.638  &  14.612  &  14.568  &  14.476  &  14.366  &  14.246  &  14.119  &  13.983  &  13.839  &  13.688   \\
 7.2  &  14.934  &  14.939  &  14.929  &  14.889  &  14.830  &  14.731  &  14.617  &  14.494  &  14.361  &  14.220  &  14.070   \\
\hline
\end{tabular}
} 
\caption {$\log\sigma_1$ for $\isotope[56]{Fe}$ at $B_{12}=1$} \label{tab:sigma0_b12_Fe}
{\tiny
\begin{tabular}{ |c||c|c|c|c|c|c|c|c|c|c|c| }
\hline
  $\log\rho$  &  $\log T$=-1.0  &  $\log T$=-0.8  &  $\log T$=-0.6  &  $\log T$=-0.4  &  $\log T$=-0.2  &  $\log T$=0.0  &  $\log T$=0.2  &  $\log T$=0.4  &  $\log T$=0.6  &  $\log T$=0.8  &  $\log T$=1.0  \\
  \hline
6.0  &  18.600  &  18.601  &  18.601  &  18.601  &  18.597  &  18.596  &  18.597  &  18.597  &  18.597  &  18.597  &  18.597   \\
 6.2  &  18.800  &  18.800  &  18.801  &  18.801  &  18.798  &  18.796  &  18.797  &  18.797  &  18.797  &  18.797  &  18.797   \\
 6.4  &  18.999  &  19.000  &  19.000  &  19.000  &  18.999  &  18.996  &  18.996  &  18.997  &  18.997  &  18.997  &  18.997   \\
 6.6  &  19.199  &  19.199  &  19.199  &  19.200  &  19.200  &  19.196  &  19.196  &  19.197  &  19.197  &  19.197  &  19.197   \\
 6.8  &  19.398  &  19.398  &  19.398  &  19.399  &  19.399  &  19.396  &  19.396  &  19.396  &  19.397  &  19.397  &  19.397   \\
 7.0  &  19.596  &  19.596  &  19.596  &  19.597  &  19.598  &  19.595  &  19.596  &  19.596  &  19.596  &  19.597  &  19.597   \\
 7.2  &  19.793  &  19.793  &  19.793  &  19.794  &  19.795  &  19.796  &  19.794  &  19.795  &  19.796  &  19.796  &  19.796   \\
\hline
\end{tabular}
} 
\end{center}
\end{table}
\begin{table}
\begin{center}
\vskip -1cm 
\caption {$\log\sigma_1$ for $\isotope[56]{Fe}$ at $B_{12}=10$} \label{tab:sigma0_b13_Fe}
{\tiny
\begin{tabular}{ |c||c|c|c|c|c|c|c|c|c|c|c| }
\hline
  $\log\rho$  &  $\log T$=-1.0  &  $\log T$=-0.8  &  $\log T$=-0.6  &  $\log T$=-0.4  &  $\log T$=-0.2  &  $\log T$=0.0  &  $\log T$=0.2  &  $\log T$=0.4  &  $\log T$=0.6  &  $\log T$=0.8  &  $\log T$=1.0  \\
  \hline
6.0  &  17.601  &  17.601  &  17.601  &  17.601  &  17.597  &  17.596  &  17.597  &  17.597  &  17.597  &  17.597  &  17.597   \\
 6.2  &  17.801  &  17.801  &  17.801  &  17.801  &  17.799  &  17.796  &  17.797  &  17.797  &  17.797  &  17.797  &  17.797   \\
 6.4  &  18.001  &  18.001  &  18.001  &  18.001  &  18.000  &  17.996  &  17.997  &  17.997  &  17.997  &  17.997  &  17.997   \\
 6.6  &  18.201  &  18.201  &  18.201  &  18.201  &  18.201  &  18.196  &  18.197  &  18.197  &  18.197  &  18.197  &  18.197   \\
 6.8  &  18.401  &  18.401  &  18.401  &  18.401  &  18.401  &  18.396  &  18.397  &  18.397  &  18.397  &  18.397  &  18.397   \\
 7.0  &  18.601  &  18.601  &  18.601  &  18.601  &  18.601  &  18.596  &  18.597  &  18.597  &  18.597  &  18.597  &  18.597   \\
 7.2  &  18.801  &  18.801  &  18.801  &  18.801  &  18.801  &  18.799  &  18.797  &  18.797  &  18.797  &  18.797  &  18.797   \\
\hline
\end{tabular}
} 
\caption {$\log\sigma_1$ for $\isotope[56]{Fe}$ at $B_{12}=100$} \label{tab:sigma0_b14_Fe}
{\tiny
\begin{tabular}{ |c||c|c|c|c|c|c|c|c|c|c|c| }
\hline
  $\log\rho$  &  $\log T$=-1.0  &  $\log T$=-0.8  &  $\log T$=-0.6  &  $\log T$=-0.4  &  $\log T$=-0.2  &  $\log T$=0.0  &  $\log T$=0.2  &  $\log T$=0.4  &  $\log T$=0.6  &  $\log T$=0.8  &  $\log T$=1.0  \\
  \hline
6.0  &  16.601  &  16.601  &  16.601  &  16.601  &  16.597  &  16.596  &  16.597  &  16.597  &  16.597  &  16.597  &  16.597   \\
 6.2  &  16.801  &  16.801  &  16.801  &  16.801  &  16.799  &  16.796  &  16.797  &  16.797  &  16.797  &  16.797  &  16.797   \\
 6.4  &  17.001  &  17.001  &  17.001  &  17.001  &  17.000  &  16.996  &  16.997  &  16.997  &  16.997  &  16.997  &  16.997   \\
 6.6  &  17.201  &  17.201  &  17.201  &  17.201  &  17.201  &  17.196  &  17.197  &  17.197  &  17.197  &  17.197  &  17.197   \\
 6.8  &  17.401  &  17.401  &  17.401  &  17.401  &  17.401  &  17.396  &  17.397  &  17.397  &  17.397  &  17.397  &  17.397   \\
 7.0  &  17.601  &  17.601  &  17.601  &  17.601  &  17.601  &  17.596  &  17.597  &  17.597  &  17.597  &  17.597  &  17.597   \\
 7.2  &  17.801  &  17.801  &  17.801  &  17.801  &  17.801  &  17.799  &  17.797  &  17.797  &  17.797  &  17.797  &  17.797   \\
\hline
\end{tabular}
} 
\caption {$\log\sigma$ for density-dependent composition} \label{tab:sigma_comp}
{\tiny
\begin{tabular}{ |c||c|c|c|c|c|c|c|c|c|c|c| }
\hline
  $\log\rho$  &  $\log T$=-1.0  &  $\log T$=-0.8  &  $\log T$=-0.6  &  $\log T$=-0.4  &  $\log T$=-0.2  &  $\log T$=0.0  &  $\log T$=0.2  &  $\log T$=0.4  &  $\log T$=0.6  &  $\log T$=0.8  &  $\log T$=1.0  \\
  \hline
6.0  &  20.144  &  20.260  &  20.376  &  20.506  &  20.628  &  20.762  &  20.902  &  21.048  &  21.200  &  21.357  &  21.522   \\
 6.2  &  20.213  &  20.310  &  20.413  &  20.537  &  20.658  &  20.785  &  20.923  &  21.066  &  21.216  &  21.372  &  21.536   \\
 6.4  &  20.293  &  20.368  &  20.456  &  20.570  &  20.689  &  20.809  &  20.944  &  21.085  &  21.233  &  21.388  &  21.550   \\
 6.6  &  20.382  &  20.435  &  20.517  &  20.607  &  20.720  &  20.834  &  20.966  &  21.104  &  21.250  &  21.403  &  21.565   \\
 6.8  &  20.477  &  20.510  &  20.572  &  20.647  &  20.753  &  20.861  &  20.989  &  21.125  &  21.268  &  21.420  &  21.580   \\
 7.0  &  20.536  &  20.555  &  20.598  &  20.671  &  20.755  &  20.857  &  20.981  &  21.114  &  21.256  &  21.406  &  21.565   \\
 7.2  &  20.633  &  20.641  &  20.666  &  20.721  &  20.792  &  20.893  &  21.006  &  21.136  &  21.276  &  21.424  &  21.581   \\
\hline
\end{tabular}
} 
\end{center}
\end{table}
\begin{table}
\begin{center}
\vskip -1cm 
\caption {$\log\sigma_0$ for density-dependent composition at $B_{12}=1$} \label{tab:sigma0_b12_comp}
{\tiny
\begin{tabular}{ |c||c|c|c|c|c|c|c|c|c|c|c| }
\hline
  $\log\rho$  &  $\log T$=-1.0  &  $\log T$=-0.8  &  $\log T$=-0.6  &  $\log T$=-0.4  &  $\log T$=-0.2  &  $\log T$=0.0  &  $\log T$=0.2  &  $\log T$=0.4  &  $\log T$=0.6  &  $\log T$=0.8  &  $\log T$=1.0  \\
  \hline
6.0  &  17.169  &  17.065  &  16.945  &  16.800  &  16.654  &  16.512  &  16.370  &  16.226  &  16.078  &  15.925  &  15.765   \\
 6.2  &  17.472  &  17.395  &  17.294  &  17.161  &  17.023  &  16.885  &  16.748  &  16.607  &  16.460  &  16.309  &  16.151   \\
 6.4  &  17.768  &  17.715  &  17.637  &  17.519  &  17.390  &  17.258  &  17.124  &  16.986  &  16.842  &  16.692  &  16.536   \\
 6.6  &  18.057  &  18.026  &  17.957  &  17.872  &  17.754  &  17.628  &  17.499  &  17.364  &  17.223  &  17.075  &  16.920   \\
 6.8  &  18.345  &  18.331  &  18.285  &  18.219  &  18.115  &  17.997  &  17.873  &  17.741  &  17.603  &  17.457  &  17.304   \\
 7.0  &  18.650  &  18.645  &  18.617  &  18.555  &  18.480  &  18.370  &  18.252  &  18.124  &  17.989  &  17.845  &  17.694   \\
 7.2  &  18.943  &  18.943  &  18.931  &  18.889  &  18.830  &  18.733  &  18.621  &  18.499  &  18.367  &  18.226  &  18.076   \\
\hline
\end{tabular}
} 
\caption {$\log\sigma_0$ for density-dependent composition at $B_{12}=10$} \label{tab:sigma0_b13_comp}
{\tiny
\begin{tabular}{ |c||c|c|c|c|c|c|c|c|c|c|c| }
\hline
  $\log\rho$  &  $\log T$=-1.0  &  $\log T$=-0.8  &  $\log T$=-0.6  &  $\log T$=-0.4  &  $\log T$=-0.2  &  $\log T$=0.0  &  $\log T$=0.2  &  $\log T$=0.4  &  $\log T$=0.6  &  $\log T$=0.8  &  $\log T$=1.0  \\
  \hline
6.0  &  15.171  &  15.068  &  14.948  &  14.801  &  14.654  &  14.512  &  14.370  &  14.226  &  14.078  &  13.925  &  13.765   \\
 6.2  &  15.475  &  15.398  &  15.297  &  15.163  &  15.024  &  14.886  &  14.748  &  14.607  &  14.461  &  14.309  &  14.151   \\
 6.4  &  15.771  &  15.718  &  15.641  &  15.521  &  15.392  &  15.259  &  15.124  &  14.986  &  14.842  &  14.692  &  14.536   \\
 6.6  &  16.061  &  16.031  &  15.961  &  15.875  &  15.757  &  15.630  &  15.500  &  15.364  &  15.223  &  15.075  &  14.920   \\
 6.8  &  16.350  &  16.336  &  16.290  &  16.224  &  16.118  &  15.999  &  15.874  &  15.742  &  15.603  &  15.457  &  15.304   \\
 7.0  &  16.657  &  16.652  &  16.625  &  16.563  &  16.486  &  16.374  &  16.254  &  16.126  &  15.990  &  15.846  &  15.694   \\
 7.2  &  16.952  &  16.954  &  16.942  &  16.901  &  16.840  &  16.740  &  16.625  &  16.501  &  16.368  &  16.227  &  16.077   \\
\hline
\end{tabular}
} 
\caption {$\log\sigma_0$ for density-dependent composition at $B_{12}=100$} \label{tab:sigma0_b13_comp}
{\tiny
\begin{tabular}{ |c||c|c|c|c|c|c|c|c|c|c|c| }
\hline
  $\log\rho$  &  $\log T$=-1.0  &  $\log T$=-0.8  &  $\log T$=-0.6  &  $\log T$=-0.4  &  $\log T$=-0.2  &  $\log T$=0.0  &  $\log T$=0.2  &  $\log T$=0.4  &  $\log T$=0.6  &  $\log T$=0.8  &  $\log T$=1.0  \\
  \hline
6.0  &  13.171  &  13.068  &  12.948  &  12.801  &  12.654  &  12.512  &  12.370  &  12.226  &  12.078  &  11.925  &  11.765   \\
 6.2  &  13.475  &  13.398  &  13.297  &  13.163  &  13.024  &  12.886  &  12.748  &  12.607  &  12.461  &  12.309  &  12.151   \\
 6.4  &  13.771  &  13.718  &  13.641  &  13.521  &  13.392  &  13.259  &  13.124  &  12.986  &  12.842  &  12.692  &  12.536   \\
 6.6  &  14.061  &  14.031  &  13.961  &  13.875  &  13.757  &  13.630  &  13.500  &  13.364  &  13.223  &  13.075  &  12.920   \\
 6.8  &  14.350  &  14.336  &  14.290  &  14.224  &  14.118  &  13.999  &  13.874  &  13.742  &  13.603  &  13.457  &  13.304   \\
 7.0  &  14.657  &  14.652  &  14.625  &  14.563  &  14.486  &  14.374  &  14.254  &  14.126  &  13.990  &  13.846  &  13.694   \\
 7.2  &  14.952  &  14.954  &  14.942  &  14.901  &  14.840  &  14.740  &  14.625  &  14.501  &  14.368  &  14.227  &  14.077   \\
\hline
\end{tabular}
} 
\end{center}
\end{table}
\begin{table}
\begin{center}
\vskip -1cm 
\caption {$\log\sigma_1$ for density-dependent composition at $B_{12}=1$} \label{tab:sigma0_b12_comp}
{\tiny
\begin{tabular}{ |c||c|c|c|c|c|c|c|c|c|c|c| }
\hline
  $\log\rho$  &  $\log T$=-1.0  &  $\log T$=-0.8  &  $\log T$=-0.6  &  $\log T$=-0.4  &  $\log T$=-0.2  &  $\log T$=0.0  &  $\log T$=0.2  &  $\log T$=0.4  &  $\log T$=0.6  &  $\log T$=0.8  &  $\log T$=1.0  \\
  \hline
6.0  &  18.600  &  18.601  &  18.601  &  18.601  &  18.597  &  18.596  &  18.597  &  18.597  &  18.597  &  18.597  &  18.597   \\
 6.2  &  18.800  &  18.800  &  18.801  &  18.801  &  18.798  &  18.796  &  18.797  &  18.797  &  18.797  &  18.797  &  18.797   \\
 6.4  &  18.999  &  19.000  &  19.000  &  19.000  &  18.999  &  18.996  &  18.996  &  18.997  &  18.997  &  18.997  &  18.997   \\
 6.6  &  19.199  &  19.199  &  19.199  &  19.200  &  19.200  &  19.196  &  19.196  &  19.197  &  19.197  &  19.197  &  19.197   \\
 6.8  &  19.398  &  19.398  &  19.398  &  19.399  &  19.399  &  19.396  &  19.396  &  19.396  &  19.397  &  19.397  &  19.397   \\
 7.0  &  19.583  &  19.583  &  19.583  &  19.585  &  19.586  &  19.582  &  19.583  &  19.584  &  19.584  &  19.584  &  19.585   \\
 7.2  &  19.780  &  19.780  &  19.780  &  19.781  &  19.783  &  19.783  &  19.782  &  19.783  &  19.784  &  19.784  &  19.784   \\
\hline
\end{tabular}
} 
\caption {$\log\sigma_1$ for density-dependent composition at $B_{12}=10$} \label{tab:sigma0_b13_comp}
{\tiny
\begin{tabular}{ |c||c|c|c|c|c|c|c|c|c|c|c| }
\hline
  $\log\rho$  &  $\log T$=-1.0  &  $\log T$=-0.8  &  $\log T$=-0.6  &  $\log T$=-0.4  &  $\log T$=-0.2  &  $\log T$=0.0  &  $\log T$=0.2  &  $\log T$=0.4  &  $\log T$=0.6  &  $\log T$=0.8  &  $\log T$=1.0  \\
  \hline
6.0  &  17.601  &  17.601  &  17.601  &  17.601  &  17.597  &  17.596  &  17.597  &  17.597  &  17.597  &  17.597  &  17.597   \\
 6.2  &  17.801  &  17.801  &  17.801  &  17.801  &  17.799  &  17.796  &  17.797  &  17.797  &  17.797  &  17.797  &  17.797   \\
 6.4  &  18.001  &  18.001  &  18.001  &  18.001  &  18.000  &  17.996  &  17.997  &  17.997  &  17.997  &  17.997  &  17.997   \\
 6.6  &  18.201  &  18.201  &  18.201  &  18.201  &  18.201  &  18.196  &  18.197  &  18.197  &  18.197  &  18.197  &  18.197   \\
 6.8  &  18.401  &  18.401  &  18.401  &  18.401  &  18.401  &  18.396  &  18.397  &  18.397  &  18.397  &  18.397  &  18.397   \\
 7.0  &  18.589  &  18.589  &  18.589  &  18.589  &  18.589  &  18.584  &  18.585  &  18.585  &  18.585  &  18.585  &  18.585   \\
 7.2  &  18.789  &  18.789  &  18.789  &  18.789  &  18.789  &  18.787  &  18.785  &  18.785  &  18.785  &  18.785  &  18.785   \\
\hline
\end{tabular}
} 
\caption {$\log\sigma_1$ for density-dependent composition at $B_{12}=100$} \label{tab:sigma0_b14_comp}
{\tiny
\begin{tabular}{ |c||c|c|c|c|c|c|c|c|c|c|c| }
\hline
  $\log\rho$  &  $\log T$=-1.0  &  $\log T$=-0.8  &  $\log T$=-0.6  &  $\log T$=-0.4  &  $\log T$=-0.2  &  $\log T$=0.0  &  $\log T$=0.2  &  $\log T$=0.4  &  $\log T$=0.6  &  $\log T$=0.8  &  $\log T$=1.0  \\
  \hline
6.0  &  16.601  &  16.601  &  16.601  &  16.601  &  16.597  &  16.596  &  16.597  &  16.597  &  16.597  &  16.597  &  16.597   \\
 6.2  &  16.801  &  16.801  &  16.801  &  16.801  &  16.799  &  16.796  &  16.797  &  16.797  &  16.797  &  16.797  &  16.797   \\
 6.4  &  17.001  &  17.001  &  17.001  &  17.001  &  17.000  &  16.996  &  16.997  &  16.997  &  16.997  &  16.997  &  16.997   \\
 6.6  &  17.201  &  17.201  &  17.201  &  17.201  &  17.201  &  17.196  &  17.197  &  17.197  &  17.197  &  17.197  &  17.197   \\
 6.8  &  17.401  &  17.401  &  17.401  &  17.401  &  17.401  &  17.396  &  17.397  &  17.397  &  17.397  &  17.397  &  17.397   \\
 7.0  &  17.589  &  17.589  &  17.589  &  17.589  &  17.589  &  17.584  &  17.585  &  17.585  &  17.585  &  17.585  &  17.585   \\
 7.2  &  17.789  &  17.789  &  17.789  &  17.789  &  17.789  &  17.787  &  17.785  &  17.785  &  17.785  &  17.785  &  17.785   \\
\hline
\end{tabular}
} 
\end{center}
\end{table}

\end{document}